# A Unified Control Framework, Mathematical Model, and Five-layer Closed-loop Case Study for Embodied Intelligence Control Systems

Yiwei Zhou[a*], Zhongcheng Lei[a], Xiaoran Dai[a], Wenshan Hu[a], Hong Zhou[a]

[a] School of Electrical Engineering and Automation, Wuhan University, Wuhan 430072, China

* *Corresponding author. Email: yiwei.zhou@whu.edu.cn*

**Abstract**

Embodied intelligence holds that an agent develops cognition and behavior through real-time body–environment interaction, so its control system must tightly couple perception, decision making, learning, optimization, constraints, and physical execution. This report asks whether a unified framework exists, how its mathematical models unify, and how the framework is realized computably.

Four conclusions emerge. (1) No single framework is accepted, but the field converges on a common skeleton: from operational-space and whole-body control to MPC, reinforcement learning, active inference, and diffusion/flow policies, all reduce to constrained sequential decision making on the configuration manifold. (2) Two languages describe it—deterministic optimal control (Euler–Lagrange plus constrained OCP, reduced in real time to a QP) and probabilistic inference (MDP/POMDP plus free-energy minimization)—rigorously bridged by control as inference and KL/path-integral control. (3) Engineering unification is already achieved by constrained optimization; the new frontier is cross-morphology, cross-task shared representation. (4) The hardest gap is provable stability and safety, the entry point for spectral, geometric, differentiable-optimization, and safe-learning methods.

A single-link manipulator then runs through all five layers: L4 parses the instruction into a reference, L3 identifies a Koopman lifted linear world model, L2 solves a receding constrained QP, L1 adds torque saturation and a CBF, and L0 is the true nonlinear plant. Closed-loop tracking RMSE is 0.0218 rad with all constraints satisfied and a world-model error near $5\times10^{-6}$ rad, reported with curves and a 3D animation.

**Keywords:** embodied intelligence; unified control framework; configuration manifold; MPC; Koopman operator; control barrier function; control as inference; world model

## 1. Introduction

Artificial intelligence is shifting from data-driven perception intelligence toward embodied intelligence centered on physical interaction [1,6]. Unlike static classification or text outputs, an embodied agent uses an actuable body to manipulate, locomote, navigate, and collaborate through a continuous perception–decision–action closed loop, facing continuous states, hard constraints, real-time demands, and unstructured, partially observable environments. Robot control has matured through classical control (PID, frequency domain), modern control (state space, optimal, adaptive, robust, Lyapunov), and the present learning-and-cognition stage that fuses data-driven methods with model-based ones [9]. These methods differ on the surface, raising the key question: do they share a unified structure, at what level, in what language, and how can it be computed?

Embodied (from embodied cognition) means an actuable body governed by dynamics and contact, multimodal perception, a real-time closed loop, and learning-based adaptation rather than full preprogramming. It does not replace traditional control: feedback control remains the low-level execution layer, with learning and cognition supplying open-environment adaptability. This report addresses three questions—framework unification (the common structure toward which the field converges), model unification (the deterministic and probabilistic languages and their bridging conditions), and example realization (a computable, reproducible five-layer instance)—and combines literature review, formal modeling, and a fixed-seed simulation, following the loop of abstract framework, mathematical unification, example realization, and comparison. Representative cognitive and architectural proposals include a unified cognitive architecture [3], distributed adaptive control [4], active inference [5], the DAO/TAO parallel view [7], and a multi-scale embodied-cognition framework [8].

The report is organized into eight chapters followed by references and methodological notes: Chapter 2 defines the five-layer architecture, Chapter 3 derives the unified mathematical skeleton, Chapter 4 compares the mathematical models and discusses representation frontiers, Chapter 5 presents the closed-loop example, Chapter 6 compares embodied intelligence with conventional control, Chapter 7 reviews applications and trends, and Chapter 8 gives topic recommendations. Throughout, peer-reviewed and real-hardware results are prioritized, preprints are marked separately, and the example is kept single-file and fixed-seed so that every quantitative claim is reproducible.

## 2. The Five-layer Architecture

"Unification" means different objects in different works—motion/force, task priorities, multi-contact, cross-morphology, or language–planning–execution. We abstract the system into five layers; every framework unifies one or a few, not all at once: L4 cognition/task, L3 representation/learning, L2 optimization/inference, L1 constraints/contact, and L0 physics/dynamics.

**Table 2-1. Five-layer genealogy of the embodied intelligence control system**

| Layer | Object of unification | Core structure | Representative unifications |
|---|---|---|---|
| L4 | Language, planning, memory, skills | Sequential modeling, skill library | Unified robot brain |
| L3 | Shared state–action representation | Latent space, equivariance, neural operators | VLA, world models, Koopman |
| L2 | How multi-task problems are solved | Constrained OCP/QP/NLP; free energy | HQP, MPC, DDP, active inference |
| L1 | Contact, actuation, safety | Holonomic constraints, CLF, CBF | Constraint-consistent inverse dynamics |
| L0 | Body–environment energy–motion laws | Configuration manifold, Euler–Lagrange/PCH | Floating-base rigid dynamics |

Classical unification focused on L0–L2 (Khatib, Righetti, Sentis, Ames); large-model unification focuses on L3–L4, such as integrated perception–cognition–execution models [2], but often weakens hard constraints and guarantees. Re-injecting physical, geometric, and safety priors into learning is the most active fusion direction. Figure 2-1 shows the central top-down five-layer backbone (green downlink), the two language panels (deterministic and probabilistic), the environment/body at the bottom, and the bottom-up feedback loop; Figure 2-2 details the backbone, with green downlink and blue dashed feedback channels.

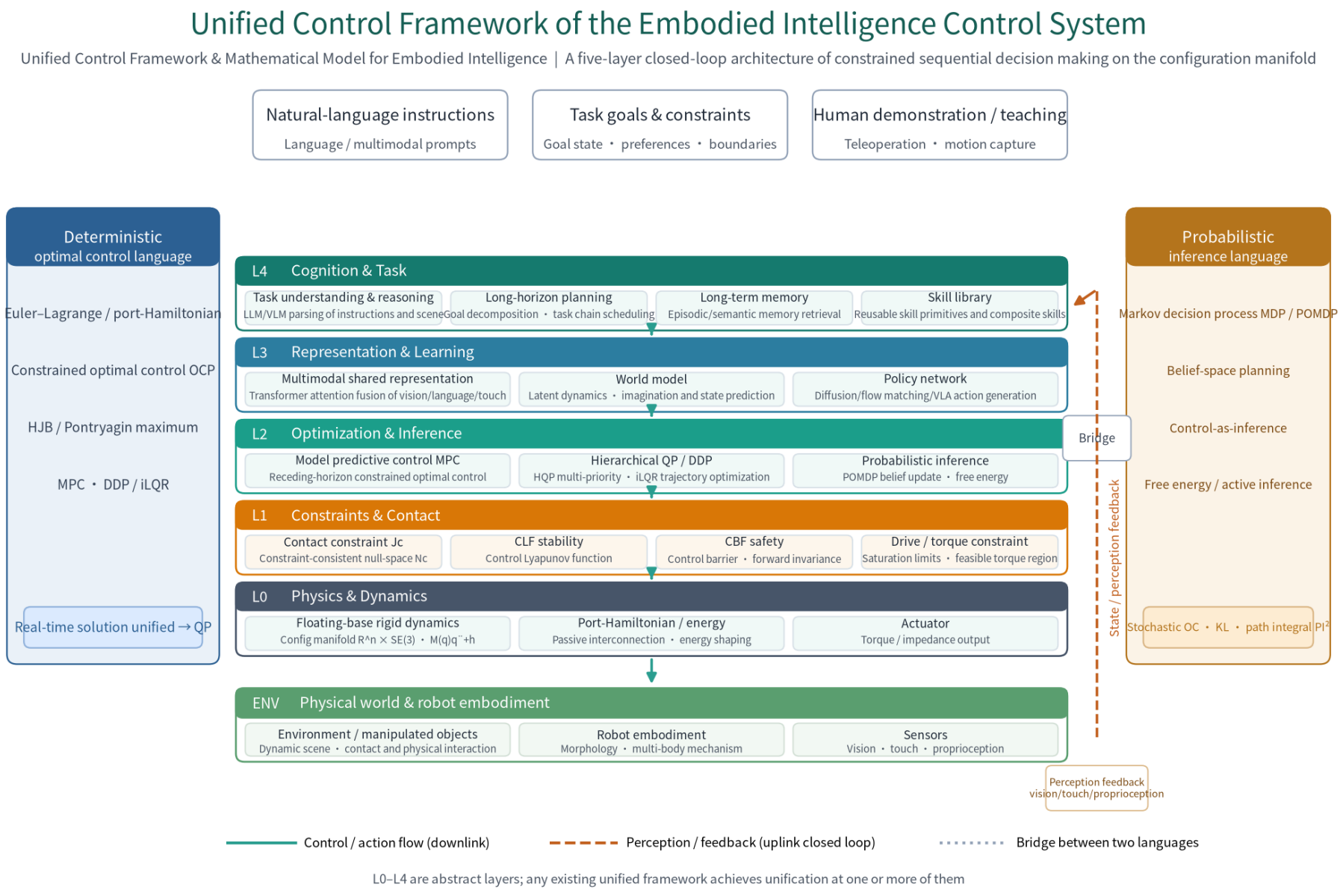

Fig. 2-1. Unified framework (five-layer backbone, two languages, feedback closed loop)

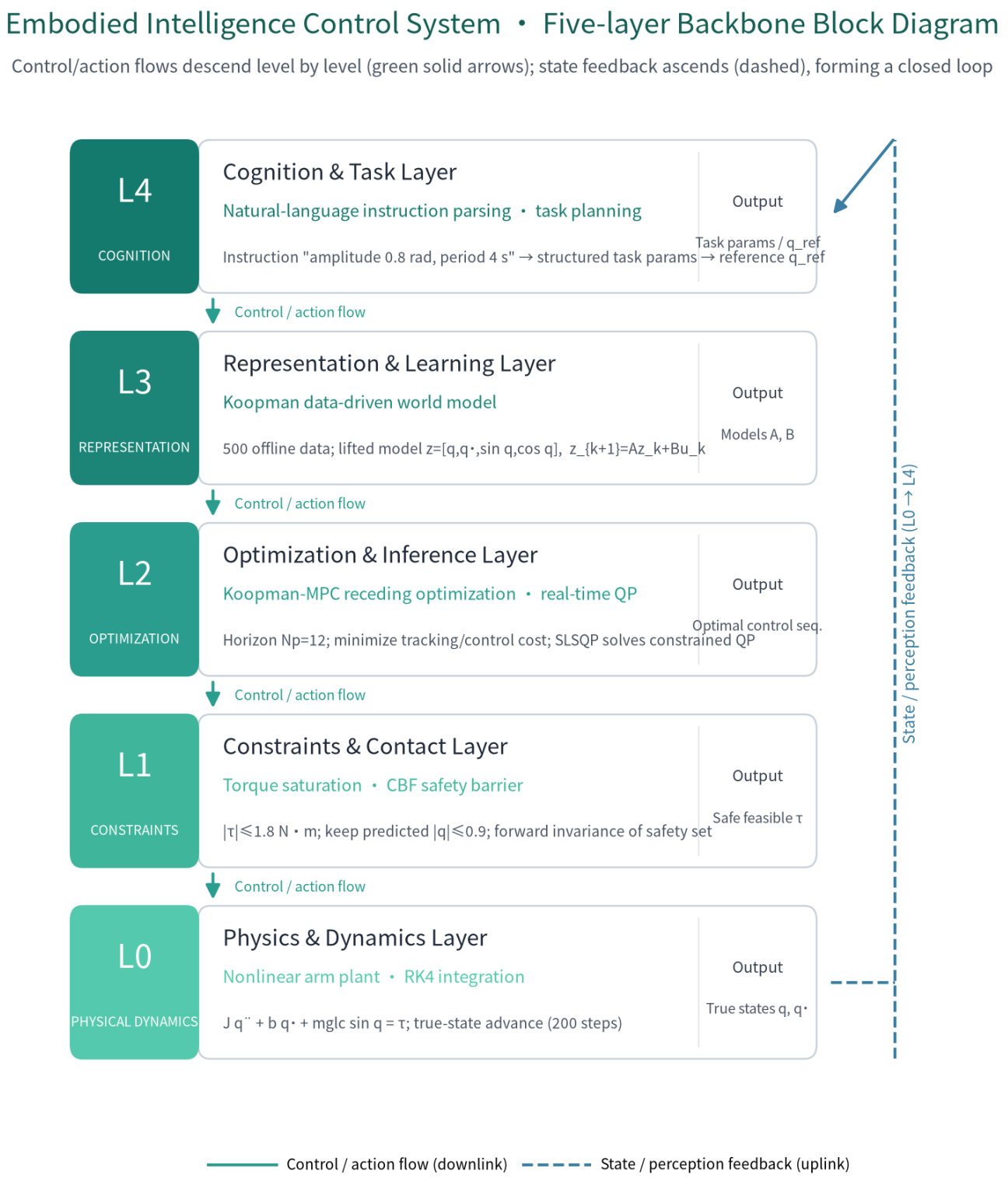

Fig. 2-2. Five-layer backbone block diagram

Standardized interfaces give composability. On the downlink, L4 passes task parameters, L3 supplies the world model and state representation, L2 gives the optimal sequence, L1 trims it to a feasible safe control, and L0 executes and returns the true state; the uplink carries states, contact, and model updates back to L2–L4.

**Table 2-2. Input–output relationships of the five-layer interfaces**

| Layer | Main inputs | Main outputs |
|---|---|---|
| L4 | Language goals, task feedback | Task parameters, reference |
| L3 | Interaction data, state | World model, representation |
| L2 | Model, reference, state | Optimal control sequence |
| L1 | Desired control, constraints | Safe feasible control |
| L0 | Control, environment | True state, observations |

The value of the five-layer structure is that it decouples the distinct responsibilities of cognition, learning, optimization, constraints, and physics: each layer can independently choose methods and be verified independently, then be combined through standardized interfaces. For example, L3 can replace the world model without changing the L2 optimizer, and L1 can add new safety constraints without changing the L0 body. This modularity also facilitates independent verification and certification procedures for each layer, which is the basis for maintainability of complex engineering systems.

The reliable engineering principle is that upper layers learn for adaptation while the bottom control ensures safety—data-driven modules handle open-environment understanding and model-based controllers with formal guarantees handle safety-critical execution. The feedback loop contains both fast low-level state feedback and slower upper-layer model updates and task adjustments, forming a nested feedback structure over multiple time scales.

## 3. The Unified Mathematical Skeleton

### 3.1 L0 physics: dynamics on the configuration manifold

Building on the operational-space formulation [10], legged/humanoid robots are underactuated floating-base systems [11,12] whose generalized coordinates combine six floating-base and $n$ joint coordinates on a configuration manifold $q = \begin{bmatrix} q_b \\ q_j \end{bmatrix} \in R^n \times SE(3), \dot{q} = (q, v) \in TQ.$, an approach deployed at scale on platforms such as Atlas [13]. Rigid-contact Euler–Lagrange equations are

$$M(q)\ddot{q} + h(q,\dot{q}) = S^{\top}\tau + J_c(q)^{\top}\lambda, \tag{1}$$

with inertia $M$, Coriolis/centrifugal/gravity $h$, actuation selection $S = [0 \vee I)$ (floating base unactuated), contact Jacobian $J_c$, and multiplier $\lambda$; contact obeys $J_c\ddot{q} + \dot{J}_c\dot{q} = 0.$. Rotation lives on the Lie group $SE(3)$ with algebra $se(3)$; geometric control gives almost-global tracking without singularities [16,17]. The same dynamics in port-Hamiltonian form $\dot{x} = [J(x) - R(x))\nabla H(x) + g(x)u, y = g(x)^{\top}\nabla H(x)$, use the Hamiltonian $H$, skew-symmetric interconnection $J$, and dissipation $R \succcurlyeq 0$, yielding passivity-based stability and compositionality [26,27]; energy tanks preserve passivity in interactive impedance and surgical control [28]. For bipedal walking, hybrid zero dynamics and virtual constraints encode stable periodic orbits [20–23].

### 3.2 L1 constraints: a unified inequality for contact, stability, safety

Inverse dynamics projects motion onto the constraint null space $N_c = I - J_c^{+J_c}$. A control Lyapunov function $V$ with $\dot{V} \leq -\alpha(V)$ encodes stability; for the safety set $C = \{x : h(x) \geq 0\}$ a control barrier function requires $\dot{h}(x,u) + \gamma h(x) \geq 0,$ [18], guaranteeing forward invariance of $C$. For a control-affine system $\dot{x} = f + gu$, both produce linear constraints on $u$, so they combine with a quadratic objective into one QP, an idea also used to make learned policies safe [19].

### 3.3 L2 deterministic language: from OCP to QP

The finite-horizon optimal control problem is

$$\min_u \sum_{k=0}^{H} \ell(x_k, u_k) + \Phi(x_H) \text{s.t.} x_{k+1} = f(x_k, u_k), \text{contact/CLF/CBF/actuation constraints.} \tag{2}$$

MPC solves it over a receding horizon [14,15]; DDP/iLQR does Bellman backpropagation [24,25]; multi-priority tasks use hierarchical QP [12]. Linearizing the continuous dynamics $\dot{x} = f(x,u)$ at the operating point $(x_0, u_0)$ gives $f(x,u) \approx f(x_0,u_0) + A_c(x - x_0) + B_c(u - u_0), A_c = \frac{\partial f}{\partial x}, B_c = \frac{\partial f}{\partial u}.$; with sampling period $T_s$ the discrete model is $x_{k+1} = A_d x_k + B_d u_k + d_k, A_d \approx I + T_s A_c, B_d \approx T_s B_c$, with bias $d_k$. For a quadratic cost $\ell(x,u) = (x - r)^{\top}Q(x - r) + u^{\top}Ru$, substituting the equalities yields a standard QP in the decision vector $u = [u_0, \ldots, u_{H-1})$:

$$\min_u \frac{1}{2}u^{\top}Hu + g^{\top}u \text{s.t.} Gu \leq h, u_{min} \leq u \leq u_{max}, \tag{3}$$

assembled from $Q, R$, with linear constraints $Gu \leq h$ carrying contact, CLF, CBF, and state limits—why model-based methods share one real-time solver.

### 3.4 L2 probabilistic language and the bridge

The MDP tuple $M = (S, A, P, r, \gamma)$, policy $\pi(a \vee s)$, and return $J = E[\sum_t \gamma^t r_t)$ obey the Bellman equation

$$V^{⬚}(s) = \max_a \left[r(s,a) + \gamma \sum_{s'} P(s' \vee s, a)V^{⬚}(s')\right), \tag{4}$$

A POMDP maintains a belief $b(s)$ updated by $b'(s') \propto p(o \vee s') \sum_s P(s' \vee s, a)b(s)$. Morphology indices as latent variables (HEAT) give a generalized POMDP [29], PSPACE-complete over a finite horizon (Dec-POMDP is NEXP-complete), explaining why cross-morphology policies need memory and amortization. Control as inference introduces an optimality variable $O_t$ with $p(O_t = 1 \vee s_t, a_t) \propto exp(r_t)$; active inference unifies perception and action via free energy [43] $F = E_q[-logp(o \vee z)) + KL[q(z) \parallel p(z)).$, with validated robot-control case studies [44], and under Gaussian/LQ assumptions it is equivalent to KL and path-integral control, though the axiomatic status of the free-energy principle remains debated.

**3.5 Preliminaries**

Lyapunov: for an equilibrium $x = 0$ a positive-definite radially unbounded $V(x)$ with $\dot{V}(x) \leq 0$ gives stability, and $\dot{V}(x) \leq -\alpha V(x)$ ($\alpha > 0$) gives exponential stability. A QP $min_u \frac{1}{2} u^\top H u + g^\top u$ under linear constraints is the real-time form of the linearized OCP. KL divergence $\boldsymbol{KL(q \parallel p) = E_q[logq - logp]}$ is the mathematical essence linking free energy and control as inference.

## 4. Model Comparison and Representation Frontiers

Table 4-1 compares 15 families. Evidence grades: A = multi-platform real hardware plus formal guarantees; B = adequate simulation, limited hardware, or mature theory; C = mainly simulation or frontier work with weak guarantees.

**Table 4-1. Mathematical model comparison of the embodied control framework families**

| Family | Object/space | Problem form | Guarantees | Grade |
|---|---|---|---|---|
| Operational-space control | Task space, Jacobian | Task-space decoupling | Force/position unification | A |
| Inverse-dynamics WBC | Constraint manifold | Constrained torque tracking | Constraint consistency | A |
| Hierarchical HQP | Priority stack | Level-by-level QP | Strict priority, real time | A |
| MPC | Receding OCP | Constrained QP/NLP | Recursive feasibility | A |
| DDP/iLQR | Bellman expansion | Local trajectory optimization | Local convergence | B |
| Geometric SE(3) | Lie group/algebra | Almost-global tracking | Singularity-free | A/B |
| Port-Hamiltonian | PCH, storage | Control as interconnection | Passivity | A/B |
| Hybrid zero dynamics | Virtual constraints | Zero-manifold orbit | Stable gait | A |
| CLF-CBF-QP | Lyapunov/barrier | Single QP | Forward invariance | A |
| Reinforcement learning | MDP | Return maximization | Conditional convergence | B |
| Active inference/FEP | Free energy | Free-energy minimization | Disputed | B/C |
| Diffusion/flow policies | SDE/ODE flow | Score/flow generation | Strong distribution, weak loop | B |
| World models | Latent dynamics | Imagination policy | Sample efficiency | B |
| Latent actions | Shared latent space | Action retargeting | Cross-embodiment | B/C |
| Koopman/neural operators | Linear lifting | Lifted linear MPC | Linear stability | B/C |

A unified brain fuses multimodal inputs in a shared Transformer and decodes world models, policies, or motion; VLA models [32,33] bind language to action and behavior foundation models [30,31] provide reusable priors, with GR00T N1 as a representative open foundation model [34]. Diffusion policies model multimodal actions via denoising and flow matching accelerates inference [37,38]; world models learn latent stochastic dynamics [35,36]; and cross-embodiment uses shared latent space, latent-action diffusion, or morphology conditioning [39,40,42], with embodiment scaling laws under study [41]. Koopman lifts nonlinear dynamics to a linear operator [45], and VLM-guided whole-body MPC extends the idea to mobile manipulation [46]. Integrated foundational models such as ACE-Brain and OmniBrain pursue the same end-to-end goal [47,48]. Open problems: hierarchy versus end-to-end, model versus data, inductive bias versus universality, the status of FEP, and the scalability of guarantees and benchmarks.

## 5. Complete Five-layer Closed-loop Example

### 5.1 Plant and L4 task

The object is a single-link rotational manipulator with gravity nonlinearity tracking a sinusoidal reference. Its dynamics are

$$J\ddot{q} + b\dot{q} + mgl_c sinq = \tau, \tag{5}$$

with joint angle $q$ (zero at hanging) and torque $\tau$; parameters $g = 9.81$, $m = 1.0$, $l = 0.5$, $l_c = 0.25$, $J = m\,l^2/3 \approx 0.0833$, $b = 0.10$, sampling period $T_s = 0.02$ (50 Hz), 200 steps over 4 s. L0 always integrates the full nonlinear model with RK4. L4 parses "swing smoothly, amplitude 0.8 rad, period 4 s" into the dictionary {motion: sin, amplitude: 0.8, period: 4.0} and generates

$$q_{ref}(t) = Asin(\omega t), \dot{q}_{ref}(t) = A\omega cos(\omega t), \tag{6}$$

with $A = 0.8$ and $\omega = 2\,\pi/4$.

**Algorithm 5-1. Instruction parsing and reference generation**

Extract amplitude and period; compute angular frequency $\omega = 2\,\pi/T$;

For each time $t$ generate reference position $q_{ref}$ and velocity $\dot{q}_{ref}$.

### 5.2 L3 Koopman world model

A fixed-seed multi-sine plus Gaussian excitation (500 points) is applied; Fig. 5-1 shows the excitation and training phase trajectory. The lifting function [45] $z = \phi(x) = [q, \dot{q}, sinq, cosq]^{\top}$, uses trigonometric features for gravity; ridge regression ($10^{-6}$) identifies, in the lifted space,

$$z_{k+1} = Az_k + Bu_k, \tag{7}$$

yielding $A \in R^{4\times4}$ and $B \in R^{4\times1}$ (Fig. 5-2): near-unit diagonals, a negative velocity-row coefficient for gravity, and an input acting only on the velocity channel.

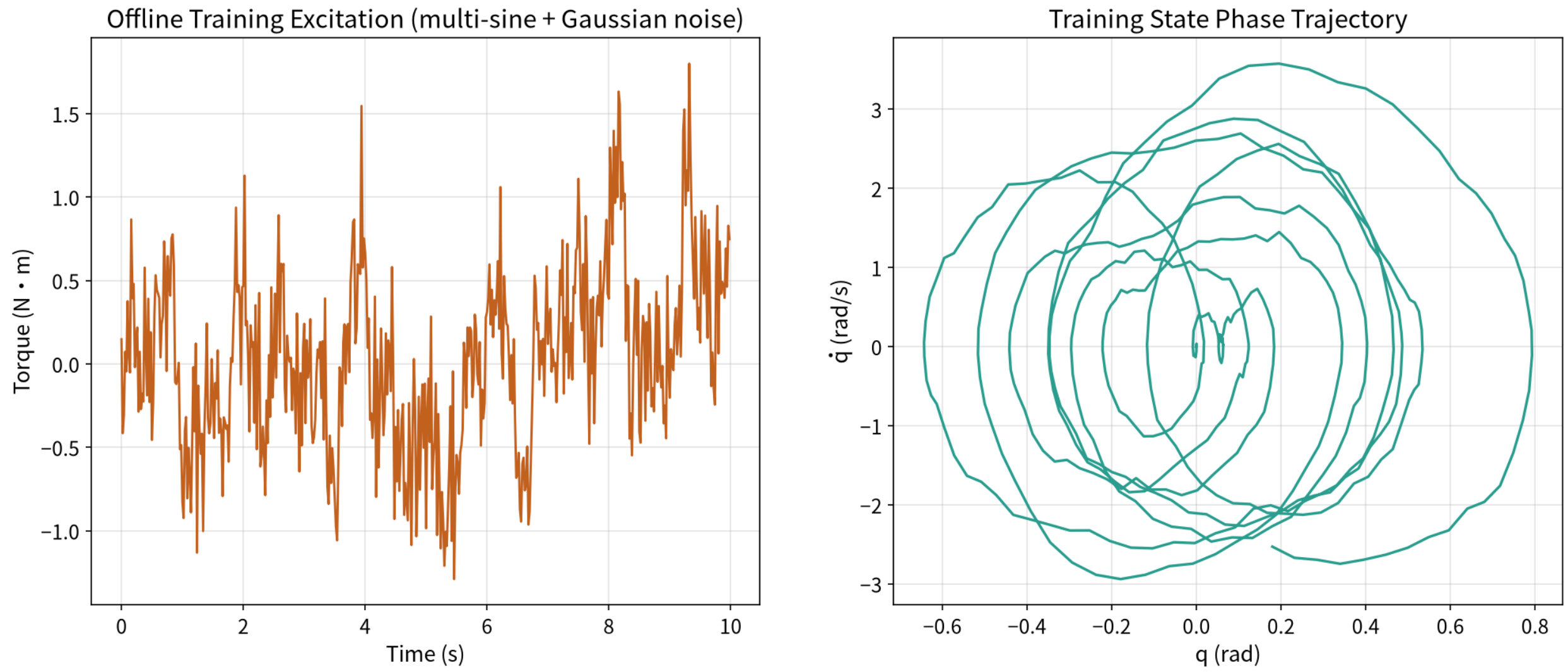

Fig. 5-1. Training excitation and training state phase trajectory

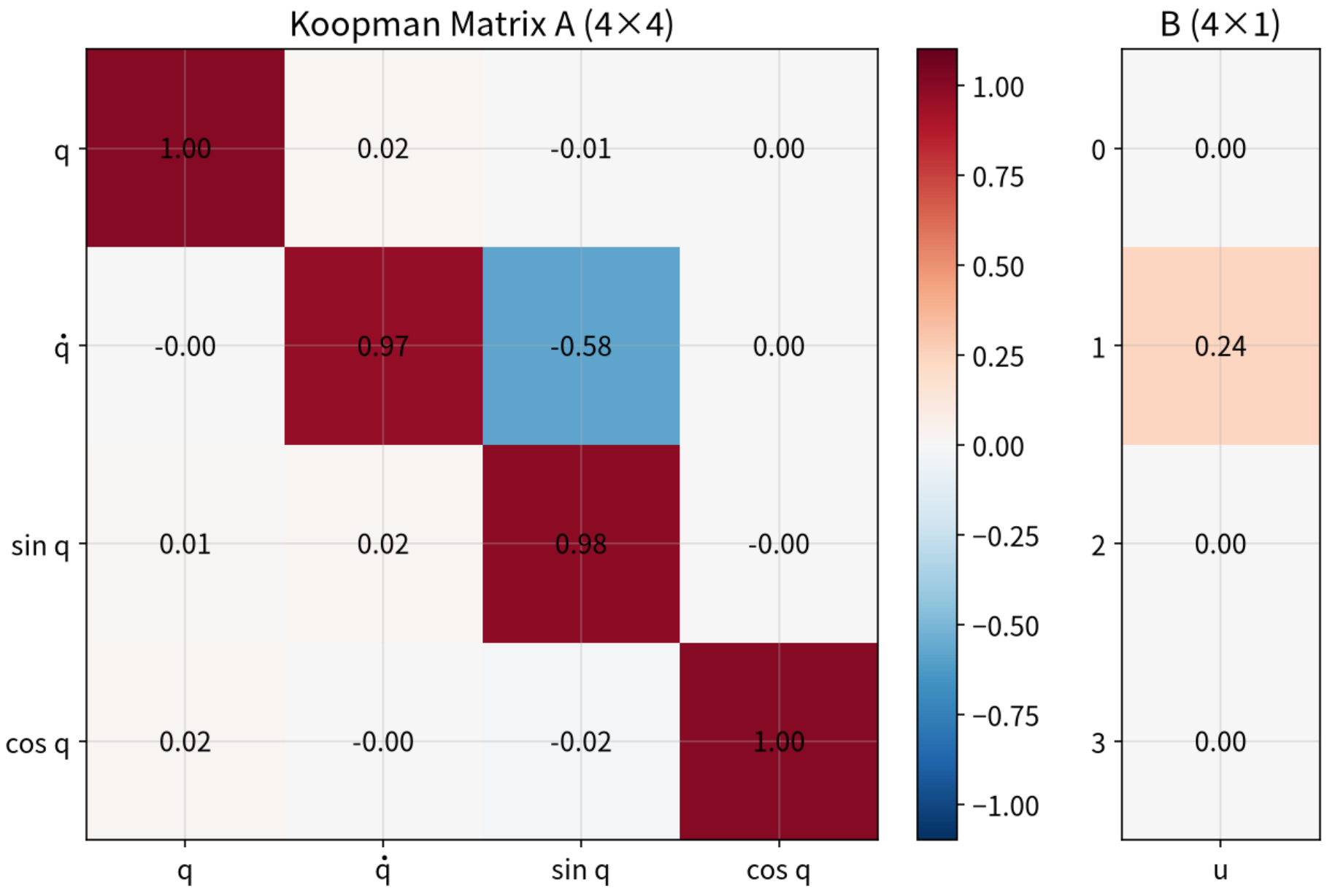

Fig. 5-2. Identified Koopman matrix A and input matrix B

**Algorithm 5-2. Koopman world-model identification**

Advance the truth with RK4, collect $(x_k, u_k, x_{k+1})$ data;

Compute features $z_k = \phi(x_k)$ and matrices $Z, Z^+, U$;

Solve ridge regression $[AB) = Z^{+[Z;U)^\top([Z;U)[Z;U)^\top+\lambda I)^{-1}}$;

A 40-step open-loop test (Fig. 5-3) tracks the truth closely, with an end error of about 8.5 mrad, confirming reliability over the MPC horizon.

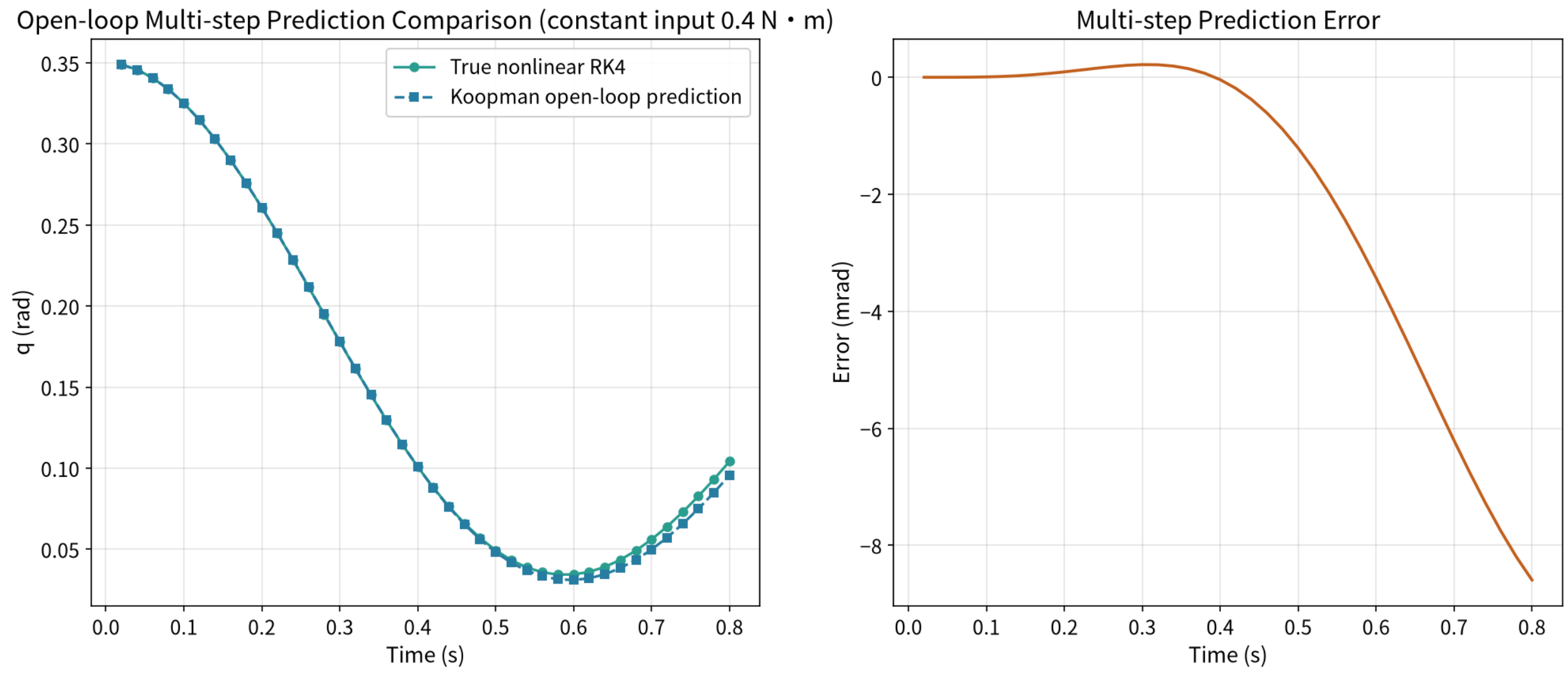

Fig. 5-3. Open-loop multi-step prediction versus truth and error

**5.3 L2 MPC, L1 constraints, and closed loop**

With horizon $N_p = 12$ and model $z_{k+1} = Az_k + Bu_k$, receding MPC minimizes

$$J = \sum_{k=1}^{N_p} \left[ W_Q (q_k - q_{ref,k})^2 + W_{QD} \dot{q}_k^2 \right) + R_U \sum_{k=0}^{N_p-1} u_k^2, \quad (8)$$

with weights $W_Q = 25$, $W_{QD} = 1$, $R_U = 0.05$; under the lifted linear model it is a constrained QP, solved with SLSQP and warm start.

**Algorithm 5-3. Receding Koopman-MPC**

Lift the current state to $z_0 = \phi(x_0)$; roll predictions along the model with decision variable $U = \left[u_0, \dots, u_{N_p-1}\right)$;

Build the quadratic cost and L1 constraints; SLSQP solves the QP; apply the first control and re-solve.

L1 imposes torque saturation

$$|\tau) \leq 1.8\text{N·m}. \tag{9}$$

and a CBF safety constraint [18] keeping the joint within the set

$$|q_k) \leq 0.9\text{rad}, \tag{10}$$

The closed loop records six columns—time $t$, reference $q_{ref}$, angle $q$, velocity $\dot{q}$, torque $\tau$, and prediction $q_{pred}$: L3 lifts the state, L2 optimizes, L1 trims, L0 advances with RK4, and the new state feeds back at 50 Hz, while identification and planning run offline.

### 5.4 Results, visualization, and comparison

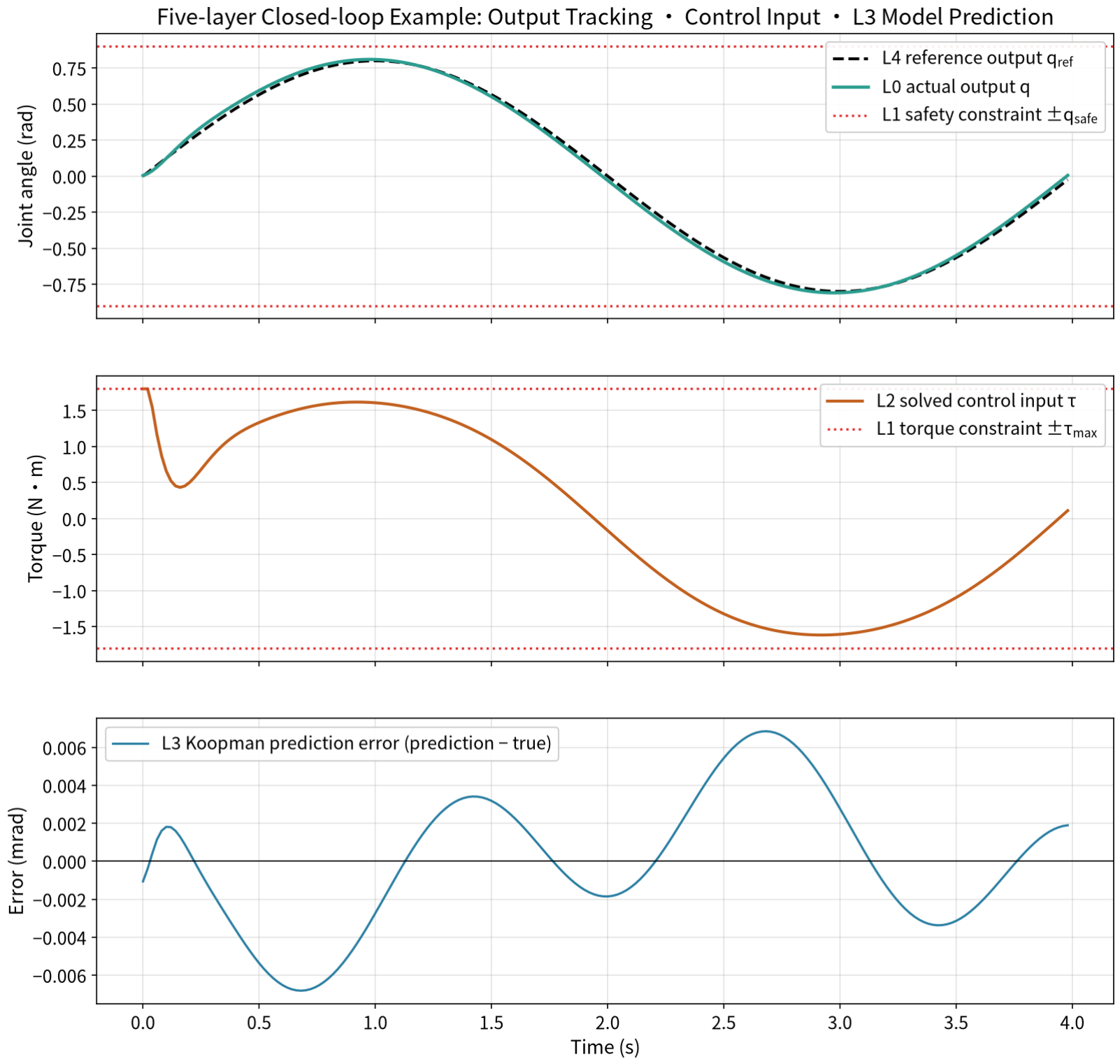

Fig. 5-4. Input–output curves (tracking, control input, L3 prediction error)

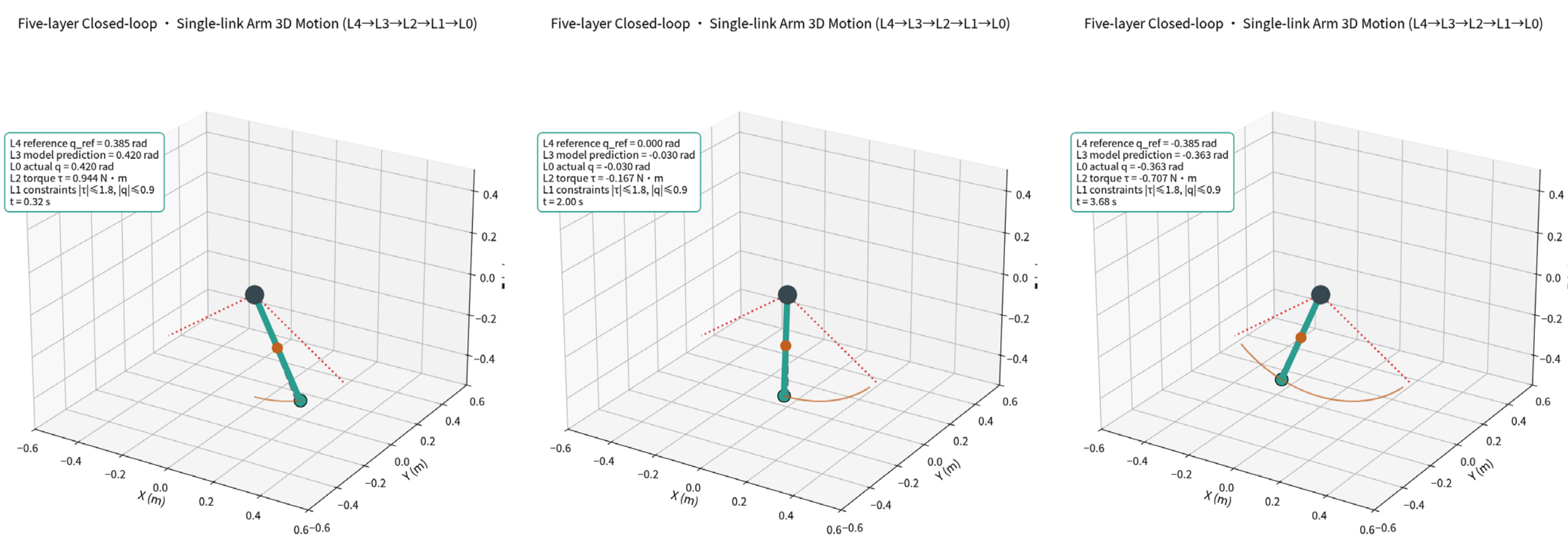

Fig. 5-5–5-7. 3D animation frames: startup, near positive amplitude, and swing-back

**Table 5-1. Quantitative results of the five-layer example**

| Metric | Value | Layer |
|---|---|---|
| World-model prediction RMSE | $5.15\times10^{-6}$ rad | L3 |
| Closed-loop tracking RMSE | 0.0218 rad | L2 |
| Maximum tracking error | 0.0344 rad | L2 |
| Torque / safety constraint | Satisfied throughout | L1 |
| Steps / duration | 200 / 4 s | L0 |

Errors are centered near zero (mean −1.44 mrad, std 21.7 mrad, no systematic bias); larger errors occur at reference zero-crossings (Fig. 5-9). Startup torque briefly hits +1.8 N·m then smooths.

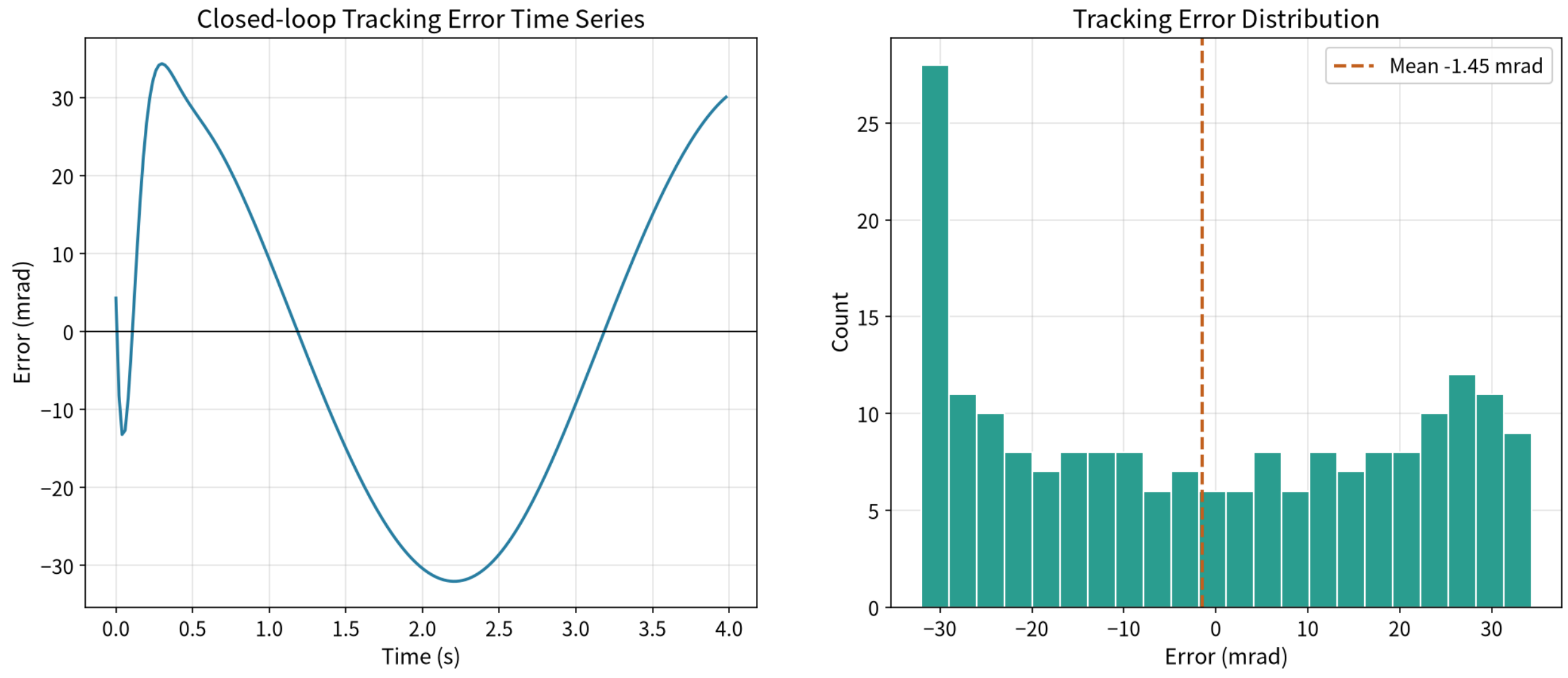

Fig. 5-9. Tracking error time series and distribution statistics

**Table 5-2. Parameter configuration**

| Module | Parameter | Value |
|---|---|---|
| Plant | mass/length/COM/inertia/friction | 1.0 / 0.5 / 0.25 / 0.0833 / 0.10 |
| L3 | points / lifting / ridge | 500 / 4 / $1\times10^{-6}$ |
| L2 | horizon / WQ / WQD / RU | 12 / 25 / 1 / 0.05 |
| L1 | torque / safety | 1.8 N·m / 0.9 rad |

The early L0–L2 example used online analytic linearization (Fig. 5-8, Table 5-3); accuracy is close, but the five-layer version adds a data-driven L3 and a language L4, giving cross-task extensibility. Error sources are finite-dimensional lifting, the horizon/weight trade-off, and saturation; limitations are the single-DOF scope, no real contact, and no online adaptation. Engineering details: SLSQP with warm start, fixed seed, and strict separation of the true plant and the model.

More specifically, the closed-loop error has three sources. First, model approximation: although the lifting includes trigonometric features, finite-dimensional lifting still truncates the true nonlinearity and multi-step open-loop prediction slowly accumulates error. Second, the horizon/weight trade-off: the finite prediction horizon and the control-energy penalty make the controller refrain from instantaneous perfect tracking in exchange for smoothness and stability. Third, actuator saturation limits the startup response. The limitations should be stated objectively: a single-link, single-DOF system does not yet involve multi-body coupling, real contact friction, or high-dimensional perception, and generalization outside the training distribution, online adaptation, and end-to-end stability proofs for the learning modules remain open. Nevertheless, the example fully demonstrates the five-layer information flow and hybrid architecture, and all results are reproducible.

Several engineering measures ensure reproducibility. A single SLSQP solver is used with constraints and bounds declared directly, avoiding extra dependencies; at each instant the shifted previous solution is used as a warm start, reducing iterations; all noise generation and data splitting use a fixed seed; and the controller always decides from the L3 approximate model while L0 is always advanced with RK4 on the full nonlinear model, the two strictly separated to avoid inflated results. Single-step optimization over the 12-step horizon completes within the 50 Hz period, and the structure extends directly to higher-DOF systems by expanding the lifting features, model dimensions, and constraint sets.

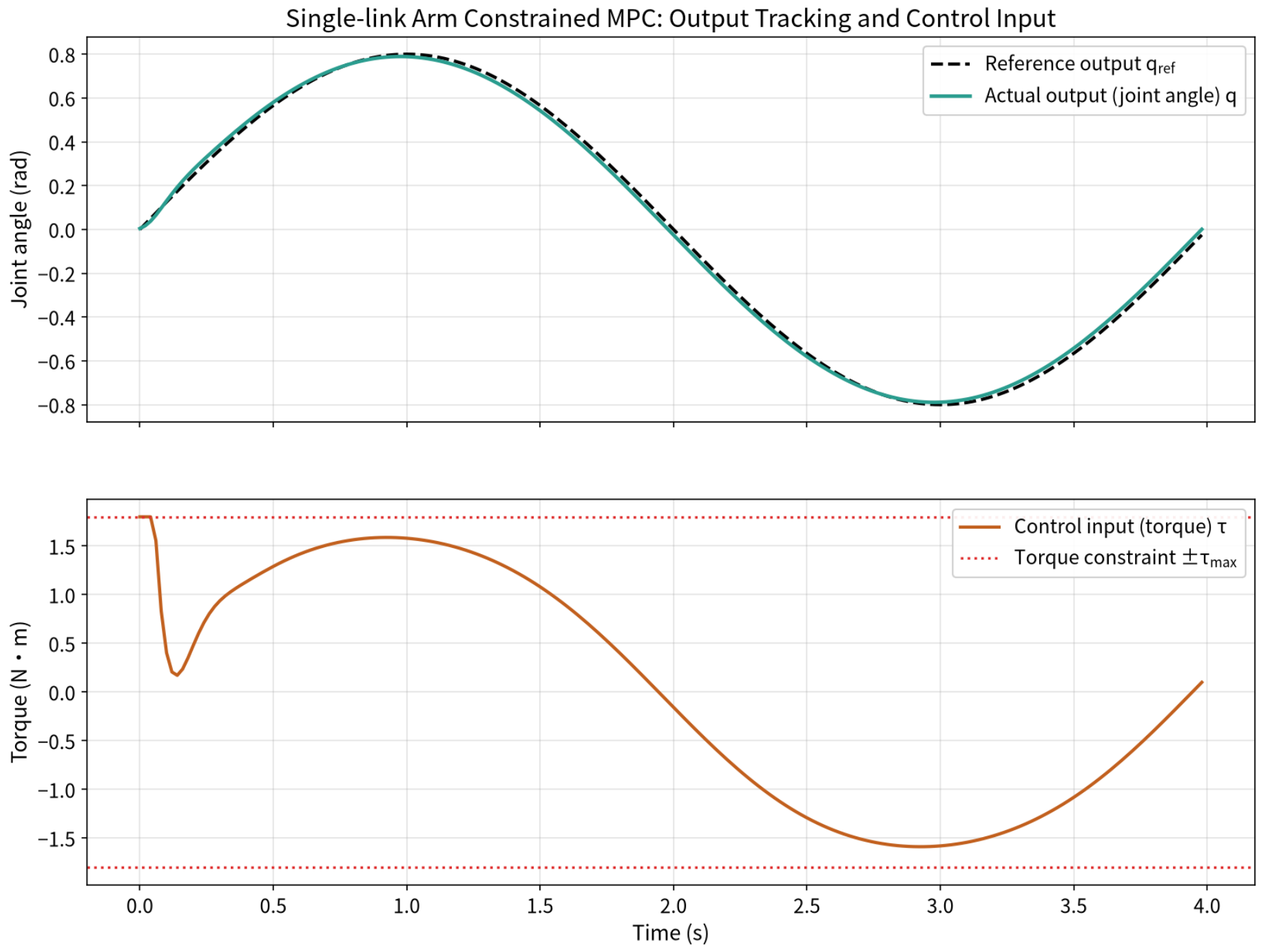

Fig. 5-8. Input–output curves of the early L0–L2 example

**Table 5-3. Early versus complete five-layer example**

| Item | Early L0–L2 | Complete five-layer |
|---|---|---|
| Task source | Pre-given reference | L4 language parsing |
| Model | Online linearization | L3 Koopman world model |
| RMSE / max error | 0.0191 / 0.0293 rad | 0.0218 / 0.0344 rad |
| Extensibility | Weak | Strong |

## 6. Embodied Intelligence versus Conventional Control

The two are hierarchical, not replacements: embodied systems use conventional control as the low-level execution layer, adding learning and cognition for open-environment adaptability at the cost of some determinism and provability.

**Table 6-1. Differences in essence and architecture**

| Dimension | Conventional control | Embodied intelligence control |
|---|---|---|
| Core idea | Feedback suppresses error | Body–environment interaction closed loop |
| Model | Transfer/state space, real-time QP | Configuration manifold, MDP/free energy, five layers |
| Environment | Structured, modelable | Unstructured, partially observable |
| Tasks | Single, predefined reference | Open, diverse, language-goal driven |
| Knowledge | Human design, offline tuning | Online/offline learning, transfer |

**Table 6-2. Capability and performance**

| Dimension | Conventional | Embodied |
|---|---|---|
| Adaptation/transfer | Limited | Strong potential |
| Accuracy | High, repeatable | Improves with data |
| Real-time | Light, deterministic | Heavy inference |
| Safety/provability | Provable, mature | Weak end-to-end |
| Data needs | Low | High |

**Table 6-3. Strengths and weaknesses**

| Class | Strengths | Weaknesses |
|---|---|---|
| Conventional | Provable, real-time, interpretable, low cost | Needs accurate models; poor generalization |
| Embodied | Adapts, transfers, handles multimodal open tasks | Hard to prove; black-box; sample-hungry |

**Table 6-4. Applications and maturity**

| Dimension | Conventional | Embodied |
|---|---|---|
| Settings | Process, servo, power, flight, joint level | Service, driving, navigation, dexterous, collaboration |
| Technologies | PID, LQR, MPC, robust, CBF | World models, RL/IL, Koopman-MPC, VLA |
| Maturity | High, standardized | Rapidly developing |

The reliable route is a hybrid: keep safety, stability, and contact in the model-based layer and perception, planning, and learning in the data-driven layer, with verifiable interfaces and hard constraints taking priority over learned desires.

A hybrid design requires three points. First, the division of responsibilities must be clear: safety, stability, and contact control that require formal guarantees stay in the model-based layer, while perceptual understanding, task planning, and representation learning in open environments go to the data-driven layer. Second, the interfaces must be verifiable: the information passed between layers should be measurable and checkable, avoiding uninterpretable implicit coupling. Third, safety constraints have the highest priority: when a learned desired action conflicts with a hard constraint, the feasible solution from the constraint layer prevails. This design both exploits the adaptivity of learning and retains the provable safety bottom line, making it the most realistic choice for current deployment.

## 7. Applications and Trends

Applications span manufacturing (flexible lines, bin picking, deformable objects, collaboration), service/home robots (language, object recognition, navigation), autonomous driving and mobile robots (sensor fusion, prediction, belief planning), legged/humanoid robots (floating-base CoM-MPC, whole-body control, learned skills), and medical/agricultural/logistics (passivity, generalization, high-tempo sorting). Key challenges are provable safety, the sim-to-real gap, real-time inference, data and evaluation, and interpretability and certification.

The challenges deserve elaboration. Provable safety is the primary deployment obstacle: when the policy is a high-dimensional neural network or generative model, traditional Lyapunov tools are hard to apply directly. The sim-to-real gap arises from dynamics differences, sensor noise, actuator delay, and contact-modeling errors, addressed through domain randomization, system identification, and online adaptation. Real-time inference demands deterministic edge latency within the sampling period; high-quality interaction data are costly and cross-embodiment benchmarks are not unified; and black-box decisions can hardly meet the audit and compliance requirements of safety-critical fields.

Trends are structured end-to-end learning with retained priors; physics-fused world models (Hamiltonian/passive/spectral constraints); cross-morphology unification and amortization; safe learning with differentiable optimization as a middle layer; and cognition–control integration. The field moves toward provable structure plus learnable residuals rather than black-box replacement or a return to pure modeling. These trends share one logic: reintroducing the mature physical structures, constraints, and guarantees of control theory into the learning-centered paradigm, so future systems move toward a middle path of provable structure and learnable residuals.

## 8. Topic Recommendations and Outlook

Prioritized directions: (1) provable stability for generative policies via Koopman/spectral lifting and CBF projection; (2) equivariant port-Hamiltonian neural dynamics on SE(3) for passive, cross-embodiment transfer; (3) strict equivalence boundaries among control as inference, FEP, and entropy-regularized RL; (4) whole-body coordination as high-dimensional second-order (including non-Hermitian) oscillator networks; (5) differentiable optimization/physics as a unified middle layer; and (6) physically consistent, stable world models.

Unification is moving from constrained optimization toward shared representation and provable safety. Short-term, safe learning and Koopman deploy on industrial and humanoid robots; medium-term, physical world models and cross-embodiment representations lower deployment cost; long-term, cognition, planning, memory, and control cooperate for long-horizon, interpretable, certifiable autonomy. High-value work sits at the intersection of provable theory and reproducible engineering. ($SE(3)$ notation appears in the geometric directions above.)

It should be soberly recognized that theoretical unification does not automatically bring engineering deployment. Progress at each level depends on repeated validation on real hardware, the establishment of standardized evaluation systems, and improved safety-certification mechanisms. In the short term, differentiable optimization, safe learning, and Koopman lifting focus on stable hybrid architectures in controlled settings; in the medium term, physical world models and cross-embodiment representations reduce long-horizon drift and the sim-to-real gap; in the long term, cognition, planning, memory, and control are expected to cooperate deeply, giving systems adaptability close to biological organisms while retaining the safety and reliability required of engineering. The truly high-value work therefore lies at the intersection of provable theory and computable, reproducible engineering.

## Appendix. Methodological Notes

**Table A-1. Main symbols and abbreviations**

| Symbol/Abbreviation | Meaning |
|---|---|
| $q$, $\dot{q}$, $\ddot{q}$ / Q, TQ | Generalized coordinates, velocity, acceleration; configuration manifold and tangent bundle |
| SE(3) / $M(q)$ / $h(q,\dot{q})$ | Special Euclidean group; inertia matrix; Coriolis/centrifugal/gravity |
| Jc / λc / V, B | Contact Jacobian; multiplier; Lyapunov/barrier functions |
| A, B / z=φ(x) / H, g | State/input matrices; lifted features; QP Hessian/gradient |
| D_KL / F | KL divergence; variational free energy |
| MPC/OCP/QP/HQP/WBC | Model predictive control; optimal control; quadratic/hierarchical QP; whole-body control |
| CLF/CBF, MDP/POMDP, VLA | Lyapunov/barrier; (partially observable) MDP; vision–language–action |
| BFM/RK4/PH/HZD/FEP | Behavior foundation model; Runge–Kutta; port-Hamiltonian; hybrid zero dynamics; free energy principle |

Notes: (1) evidence is graded from systematic retrieval; arXiv/preprint items lack full peer review and should be rechecked on publication. (2) Verify volumes/pages by DOI and proceedings. (3) The Chapter 5 simulation is single-file, fixed-seed (np.random.seed(42)), and directly computed; the 3D animation is a full video represented by frames. (4) The report does not replace real-hardware experimental design and validation.